\documentclass{article}
\usepackage{graphicx}

\begin{document}

\title{Hysteretic method for measuring the flux trapped within the core of a superconducting lead-coated ferromagnetic torus by a linked superconducting tin ring, in a novel Aharonov-Bohm-like effect based on the Feynman path-integral principle}
\date{White paper of May 26, 2012 by R. Y. Chiao \cite{Chiao2011}}
\author{}
\maketitle
\abstract{A novel kind of nonlocal, macroscopic Aharonov-Bohm effect involving two topologically linked superconducting rings made out of two different materials, namely, lead and tin, is suggested for experimental observation, in which the lead ring is a torus containing a core composed of permanently magnetized ferromagnetic material. It is predicted that the remnant fields in a hysteresis loop induced by the application of a magnetic field imposed by a large external pair of Helmholtz coils upon the tin ring, will be asymmetric with respect to the origin of the loop. An appendix based on Feynman's path-integral principle is the basis for these predictions.}
\pagebreak
\begin{figure}
\includegraphics[width=4.75in]{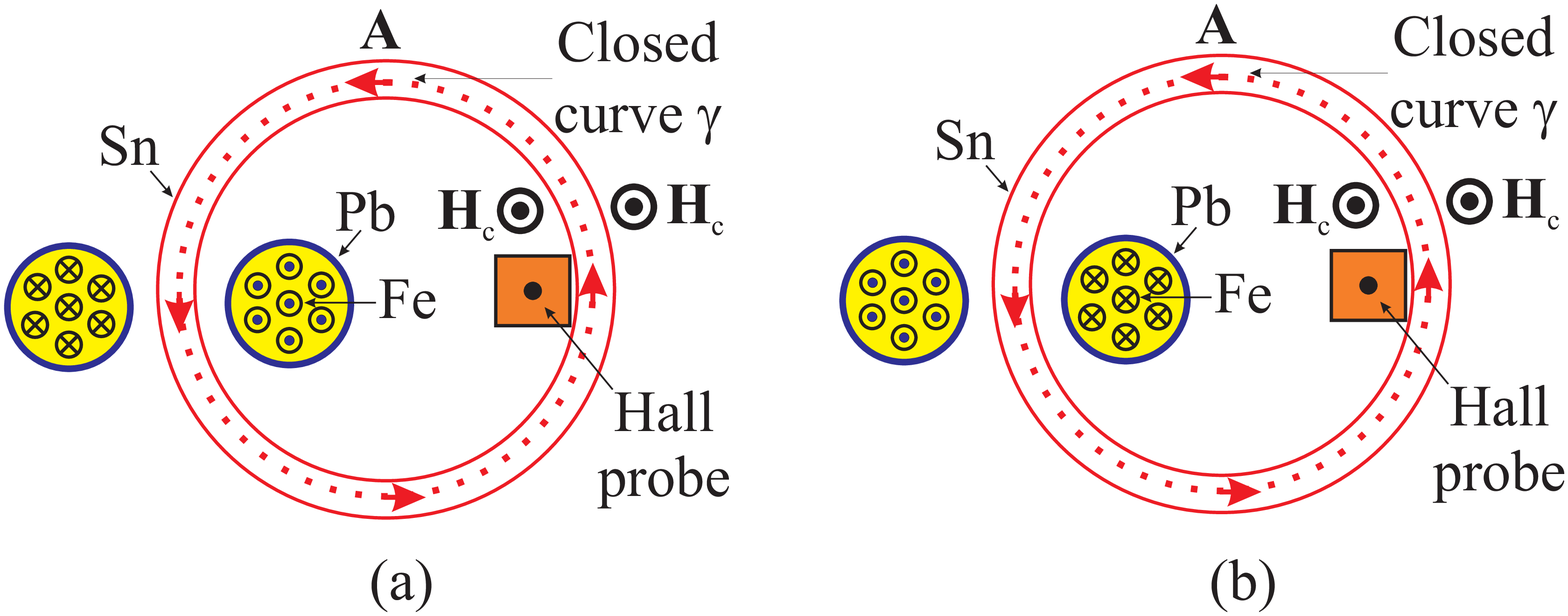}
\caption{(a) Two topologically-linked
superconducting (SC) rings are made out of two different SC materials, tin
(Sn) and lead (Pb). The lead ring (i.e., the torus shown in a cross
sectional view with a blue rim, and with the black dots and crosses denoting
trapped ferromagnetic flux filling its interior) has the higher transition
temperature and field, but the tin ring (the circle in red) has the lower
transition temperature and field. The lead torus (outlined in blue) encloses
within it a magnetized ferromagnetic (Fe) core (filled in with yellow), and
traps within its interior a field of around 200 Gauss, which points in the 
\emph{same} direction as the externally applied critical field $\bf{H}_{%
\rm{c}}$ of tin. (b) Same as (a), except that the torus traps within its
ferromagnetic core a field pointing in the \emph{opposite} direction to the
applied critical field $\bf{H}_{\rm{c}}$ of tin. The Hall probe
measures the remnant field trapped within the tin ring at the zero-crossings
of the applied magnetic field $\bf{H}$ for both (a) and (b). $\bf{A}$
denotes the \emph{total} vector potential evaluated at the dashed circle from
$all$ sources.}
\end{figure}

In Bong-Soo Kang's
experiment, in addition to the observation of the angular momentum
associated with the vector potential, there is yet another interesting
possibility for observing and measuring another kind of Aharonov-Bohm-like
effect (see Figure 1). If the tin ring in Figure 1 is much thicker than a
penetration depth, then the quantization of flux trapped inside this ring
will obey the following condition:%
\begin{equation}
\Phi _{n}=\oint\limits_{\gamma }{\bf A}\cdot d{\bf l}=n\frac{h}{2e},%
n=0,\pm 1,\pm 2,...  \label{flux quantization}
\end{equation}%
where $\Phi _{n}$ is the trapped flux, $\bf{A}$ is the \emph{total}
vector potential arising from $all$ sources evaluated along the closed curve $%
\gamma $ deep within the material of the tin ring (e.g., the dashed circle $%
\gamma $ in Figure 1), $d\bf{l}$ is a line element of this closed curve $%
\gamma $, $n$ is an integer, $h$ is Planck's constant, and $e$ is the
electron charge.

In the space enclosed by the tin ring in Figure 1, there is a
superconducting lead-coated torus containing a ferromagnetic core. This kind
of topological configuration, which has a nontrivial topological linking
number of unity, is isomorphic to that of the ferromagnetic torus overcoated
by superconducting niobium used by Tonomura et al., by which they
convincingly demonstrated the Aharonov-Bohm effect \cite{Tonomura}. The
superconducting lead coating of the torus will quantize the magnetic flux
contained within the ferromagnetic core of the torus. However, more
importantly, the magnetic field arising from the ferromagnetic material
within the torus will be completely shielded by the Meissner effect due to
the thick overcoating of the superconducting lead \cite{Decker}, and will thereby
be prevented from leaking into the space exterior to the torus in which the
tin ring resides. Hence from a purely classical viewpoint, the electrons
within the tin ring can feel no forces arising from the ferromagnetic field
which is entirely confined to the interior of the superconducting lead
torus. Therefore, classically, there could exist no influence whatsoever of
the ferromagnetic field \emph{interior} to the lead torus upon the tin ring \emph{%
exterior} to the lead torus. If any such influence were to be observed in our
experiments, it would imply a violation of the principle of \emph{locality}
implicit in classical physics. A positive result in our experiments would
constitute evidence for a new kind of macroscopic quantum \emph{nonlocality}%
, which arises from a Aharonov-Bohm-type quantum holonomy that is entirely
nonclassical in origin and nature (see Appendix).

At the quantum level of description, there will arise a contribution of the
vector potential along the dashed circle $\gamma $ in Figure 1 due to the
trapped ferromagnetic flux contained within the torus, which $\gamma $
encloses, to the accumulated quantum mechanical phase of the Cooper pair
wavefunction. This phase shift arises from the Aharonov-Bohm effect. 

Now consider a hysteresis-loop procedure in which we apply an external
magnetic field from a large pair of Helmholtz coils upon the entire assembly
shown in Figure 1, starting initially from zero field. Let us assume that
the ambient temperature is close to absolute zero, and that there is
initially no magnetic field trapped by the tin ring. Then when the applied
external magnetic field reaches the critical field of tin \cite{Shaw}, i.e.,%
\begin{equation}
H_{\rm{c}}\approx 280\rm{ Gauss}
\end{equation}%
the tin ring will be driven into the normal state, and the externally
applied magnetic field will begin to fill the space enclosed by the tin ring
(except for the area already filled by the superconducting lead-coated
ferromagnetic torus), so that there will now be two contributions to the
vector potential%
\begin{equation}
\bf{A}=\bf{A}_{\rm{1}}+\bf{A}_{\rm{2}}
\end{equation}%
where $\bf{A}_{\rm{1}}$ is the contribution arising from the flux
contained within the ferromagnetic core of the superconducting lead-coated
torus, and $\bf{A}_{\rm{2}}$ is the contribution from the flux that
now uniformly fills the rest of the space enclosed by the tin ring due to
the externally applied magnetic field, whose magnitude will slightly exceed
the critical field $H_{\rm{c}}$ of the tin material of the ring.

If the externally applied magnetic field from the pair of Helmholtz coils is
subsequently reduced down below $H_{\rm{c}}$, then the tin ring will once
again become superconducting, and therefore the tin ring will trap a certain
number of quantized flux lines given by $n_{\rm{remnant}}$ where 
\begin{equation}
\left. \Phi _{n}\right\vert _{\rm{trapped}}=\oint\limits_{\gamma }\left( 
{\bf A}_{\rm{1}}+{\bf A}_{\rm{2}}\right) \cdot d {\bf l}=n_{%
\rm{remnant}}\frac{h}{2e}  \label{n_remnant}
\end{equation}%
If the externally applied magnetic field from the Helmholtz coils is now
further reduced down to zero, the integer $n_{\rm{remnant}}$ will still
remain a constant, and therefore there will still remain a remnant flux
trapped within the tin ring corresponding to the constant integer $n_{\rm{%
remnant}}$, which can then be measured by the Hall probe in Figure 1 at the
zero-crossing of the applied magnetic field.

Note that the integer $n_{\rm{remnant}}$ will depend on the relative
direction of the ferromagnetic field trapped in the \emph{interior} of the
lead-coated torus, relative to the direction of the \emph{exterior}, applied
critical field $H_{\rm{c}}$, as illustrated in the two cases (a) and (b)
in Figure 1. In other words, the integer $n_{\rm{remnant}}$ in (\ref%
{n_remnant}) will depend on whether the relative contributions from $\bf{%
A}_{\rm{1}}$ and $\bf{A}_{\rm{2}}$ will reinforce each other, or
whether they will tend to cancel each other out. This should result in an 
\emph{asymmetric} hysteresis loop with \emph{asymmetric} positive and
negative remnant fields, as the externally applied magnetic field from the
Helmholtz coils is cycled periodically, and symmetrically, from $+H^{\prime
} $ to $-H^{\prime }$, where $\left\vert H^{\prime }\right\vert $ slightly
exceeds $H_{\rm{c}}$.

\section*{Appendix: Feynman path-integral principle}

The Feynman path-integral principle begins with the fact that for any
particle, e.g., the electron, propagating through spacetime%
\begin{equation}
d\phi \propto dS  \label{phase-action-relation}
\end{equation}%
where, when the electron is viewed as being a \emph{propagating wave}, $%
d\phi $ is the infinitesimal \emph{phase} acquired by the electron due to
its propagation through spacetime along an infinitesimal spacetime path
element $dx^{\mu }$ ($\mu =0,1,2,3$), and where, when the electron is viewed
as being a \emph{moving particle}, $dS$ is the infinitesimal \emph{action}
acquired by the same electron due to its motion along the same infinitesimal
spacetime path element $dx^{\mu }$. Thus the relationship $d\phi \propto dS$
is an expression of the \emph{wave-particle duality} of the electron, with $%
d\phi $ being a \emph{wave} property of the electron, and $dS$ being a \emph{%
particle} property of the electron.

The proportionality constant in the relationship $d\phi \propto dS$ must be
determined by experiment, and turns out to be $1/\hbar $, where $\hbar $ is
the reduced Planck's constant, so that (\ref{phase-action-relation}) becomes%
\begin{equation}
d\phi =\frac{1}{\hbar }dS  \label{phase-action-with-hbar}
\end{equation}%
For the special case of a free electron propagating through the vacuum in
the absence of any external forces%
\begin{equation}
d\phi =k_{\mu }dx^{\mu }
\end{equation}%
where $k_{\mu }$ is the four-wavevector of the free electron's plane-wave
solution (i.e., the solution of the Dirac equation in a field-free vacuum),
and where $dx^{\mu }$ is an infinitesimal four-displacement, i.e., an
infinitesimal spacetime path element along which the free electron is
traveling. It then follows that%
\begin{equation}
d\phi =\frac{1}{\hbar }dS=k_{\mu }dx^{\mu }
\end{equation}%
and therefore that%
\begin{equation}
dS=\hbar k_{\mu }dx^{\mu }=p_{\mu }dx^{\mu }
\end{equation}%
where%
\begin{equation}
p_{\mu }=\hbar k_{\mu }  \label{four-momentum-four-wavevector-relation}
\end{equation}%
is the four-momentum of the electron. From (\ref%
{four-momentum-four-wavevector-relation}), it follows that%
\begin{equation}
{\bf p}=\hbar {\bf k}  \label{DeBroglie}
\end{equation}%
\begin{equation}
E=\hbar \omega   \label{Einstein-Planck}
\end{equation}%
where $\bf{p}$ is the three-momentum of the electron, and $\bf{k}$
is its three-wavevector, and where $E$ is the energy of the electron, and $%
\omega $ is its angular frequency. The first of these two relationships, (%
\ref{DeBroglie}), is the De Broglie law for the electron, and the second of
these two relationships, (\ref{Einstein-Planck}), is the Planck-Einstein
relationship, which was first discovered in connection with the photon, but
also applies to the electron.

Note that the above phase-action relationship (\ref{phase-action-with-hbar})
would apply to any neutral particle, as well as to the charged electron. Is
there any extra contribution to the phase or the action arising from the
fact that the electron is charged? Experimentally, we know that the answer
has to be yes to this question, because we know that the electron interacts
with electromagnetic fields via its charge $q=e$. Hence we must add to the
action (\ref{phase-action-relation}) an extra piece that describes the
interaction of the charge $q$ of a charged particle with an externally
applied electromagnetic field, i.e.,%
\begin{equation}
dS_{q}\propto q  \label{S linear in q}
\end{equation}%
since we expect the size of the charge-field interaction to be proportional
to the size of the charge $q$. Next, we seek an action $dS_{q}$ which
satisfies the \emph{linearity} requirement%
\begin{equation}
dS_{q}\propto dx^{\mu }  \label{S linear in dx}
\end{equation}%
for infinitesimal four-displacements $dx^{\mu }$ of the charge $q$ through
spacetime in the presence of electromagnetic fields. This requirement
follows from the fact that all physically reasonable\ fields must become 
\emph{uniform} fields when viewed on the tiny, infinitesimal length scales
given by $dx^{\mu }$. Hence the action of displacing a charge in the
presence of such uniform fields by an infinitesimal amount $dx^{\mu }$ must
be \emph{linear} in $dx^{\mu }$. Putting (\ref{S linear in q}) and (\ref{S
linear in dx}) together, we therefore get%
\begin{equation}
dS_{q}\propto qdx^{\mu }
\end{equation}

However, note that, whereas $dS_{q}$ is a four-scalar and the charge $q$ is
also a four-scalar, $dx^{\mu }$ is, by contrast, a four-vector. Following
the spacetime symmetry argument given by Landau and Lifshitz \cite{LXL}, one
must therefore contract the contravariant four-vector $dx^{\mu }$ with some
covariant four-vector $A_{\mu }$ in order to be able to form the invariant
four-scalar%
\begin{equation}
dS_{q}\propto qA_{\mu }dx^{\mu }  \label{dS_q}
\end{equation}%
The only covariant four-vector that can satisfy Landau and Lifshitz's
symmetry requirement, and also yield the correct non-relativistic limit \cite%
{determination of vector potential}, is the \emph{electromagnetic}
four-vector potential $A_{\mu }$. 

By taking the non-relativistic limit, and by obtaining the non-relativistic
Lorentz force law from this limit \cite{Chiao2011}, one also uniquely
determines the sign and the proportionality constant of (\ref{dS_q}). One
then finds that in SI\ units \cite{Chiao2011}%
\begin{equation}
dS_{q}=+qA_{\mu }dx^{\mu }  \label{dS_q=+qA_u.dx^u}
\end{equation}%
This argument also uniquely determines the minimal coupling rule for quantum
mechanics \cite{Chiao2011}.

Let us now integrate (\ref{dS_q=+qA_u.dx^u}) along some specified spacetime
path from event $a$ to event $b$%
\begin{equation}
\left. S_{q}\left[ \rm{path}\right] \right\vert _{a}^{b}=\left. \int
qA_{\mu }dx^{\mu }\left[ \rm{path}\right] \right\vert _{a}^{b}
\label{path integral for dS_q}
\end{equation}%
The phase accumulated by the charge $q$ along this path from $a$ to $b$ will
be given by%
\begin{equation}
\left. \phi _{q}\left[ \rm{path}\right] \right\vert _{a}^{b}=\frac{1}{%
\hbar }\left. S_{q}\left[ \rm{path}\right] \right\vert _{a}^{b}=\frac{q}{%
\hbar }\left. \int A_{\mu }dx^{\mu }\left[ \rm{path}\right] \right\vert
_{a}^{b}  \label{phase along specified path}
\end{equation}%
\qquad 

At this point, it may be objected that it is always possible to choose a
gauge\ such that the accumulated phase $\left. \phi _{q}\left[ \rm{path}%
\right] \right\vert _{a}^{b}$ becomes identically zero, because one could in
principle always arbitrarily choose a scalar function $\chi $ such that%
\begin{equation}
A_{\mu }\rightarrow A_{\mu }^{\prime }=A_{\mu }+\partial _{\mu }\chi =0
\end{equation}%
\emph{at each point} along the specified path between $a$ and $b$ in (\ref%
{phase along specified path}). 

However, when there is a \emph{closed} path $\gamma $ which \emph{encloses}
a nonzero amount of magnetic flux, such as in the configuration illustrated
in Figure 1, there arises a \emph{quantum holonomy} which prevents this from
happening, viz.,%
\begin{equation}
\phi _{q}\left[ \gamma \right] =\frac{1}{\hbar }S_{q}%
\left[ \gamma \right] =\frac{q}{\hbar }%
\oint\limits_{\gamma }A_{\mu }dx^{\mu }\neq 0
\end{equation}%
From Stokes's theorem, one finds that%
\begin{equation}
\phi _{q}\left[ \gamma \right] =\frac{q}{\hbar }%
\oint\limits_{\gamma }A_{\mu }dx^{\mu }=\frac{q}{\hbar }\Phi _{\gamma }\neq 0
\end{equation}%
where the magnetic flux $\Phi _{\gamma }$ enclosed by the closed curve $%
\gamma $ is given by%
\begin{equation}
\Phi _{\gamma }=\oint\limits_{\gamma }A_{\mu }dx^{\mu }\neq 0
\end{equation}
which is a gauge-invariant quantity.

If one further chooses the closed curve $\gamma $ to lie deep inside the
superconducting material of the tin ring, such as the dashed red circle of
Figure 1, and if one sets $q=2e$ to be the charge of the Cooper pairs deep
inside the superconductor, then it follows from the \emph{single-valuedness}
of the Ginzburg-Landau superconducting order parameter that the magnetic
flux $\Phi _{\gamma }$\ enclosed by $\gamma $\ must be quantized in integer
units of the fundamental flux quantum, i.e., 
\begin{equation}
\Phi _{\gamma }=\oint\limits_{\gamma }{\bf A}\cdot d{\bf l}=n\frac{h}{%
2e}
\end{equation}%
where $n$ is an integer, and therefore one recovers the flux quantization condition (\ref{flux
quantization}). The rest of the above argument in the main body of the paper for a novel, macroscopic manifestation of the Aharonov-Bohm effect then follows.

\pagebreak

\end{document}